\documentclass{article}
\usepackage{spconf,amsmath,graphicx,bbm}
\usepackage{algorithm}
\usepackage{algorithmic}
\usepackage{tabularx}

\newcolumntype{L}[1]{>{\raggedright\arraybackslash}m{#1}}
\newcolumntype{C}[1]{>{\centering\arraybackslash}m{#1}}
\newcolumntype{R}[1]{>{\raggedleft\arraybackslash}m{#1}}

\title{L-Vector: Neural Label Embedding for Domain Adaptation}
\name{Zhong Meng$^{1}$, Hu Hu$^{1,2}$\sthanks{Work performed during an internship
	at Microsoft.}, Jinyu Li$^{1}$, Changliang Liu$^{1}$, Yan
Huang$^{1}$, Yifan Gong$^{1}$,Chin-Hui Lee$^{2}$}
\address{$^{1}$ Microsoft Corporation, Redmond, WA, USA \\ $^{2}$ Georgia Institute of Technology, Atlanta, GA, USA}
\begin{document}
\ninept
\maketitle
\begin{abstract}
We propose a novel neural label embedding (NLE) scheme for the domain
adaptation of a deep neural network (DNN) acoustic model with
\emph{unpaired} data samples from source and target domains.  With NLE method, we
distill the knowledge from a powerful source-domain DNN into a
dictionary of label embeddings, or \emph{l}-vectors, one for each senone class. 
Each \emph{l}-vector is a representation of the senone-specific output 
distributions of the source-domain DNN and is learned to minimize the
average $L_2$, Kullback-Leibler (KL) or symmetric KL distance to the output vectors with the same label through simple averaging or standard back-propagation. 
During adaptation, the \emph{l}-vectors serve as the soft targets to train
the target-domain model with cross-entropy loss.
Without parallel data constraint as in the teacher-student learning, NLE is specially suited for the situation where the paired target-domain data
cannot be simulated from the source-domain data. We adapt a 6400 hours
multi-conditional US English acoustic model to each of the 9 accented English (80 to
830 hours) and kids' speech (80 hours). NLE achieves up to
14.1\% relative word error rate reduction over direct re-training with one-hot labels. 
\end{abstract}


\section{Introduction}
\label{sec:intro}
Deep neural networks (DNNs) \cite{DNN4ASR-hinton2012, sainath2011making, deng2013recent} have greatly advanced the performance of automatic speech recognition (ASR) with a large amount of training data. However, the performance degrades when test data is from
a new domain. Many DNN adaptation approaches were proposed to compensate for the acoustic mismatch between training and testing.
In \cite{kld_yu, map_huang, l2_liao}, regularization-based approaches restrict the neuron output distributions or the model parameters to stay not too far away from the source-domain model. 
In \cite{lhn, feature_seide}, transformation-based approaches reduce the number of learnable parameters by updating only the transform-related parameters. In \cite{svd_xue_1,svd_xue_2}, the trainable parameters are further reduced by singular value decomposition of weight matrices of a neural network. 
In addition, i-vector \cite{ivector_saon} and speaker-code \cite{sc_abdel, sc_xue} are used as auxiliary features to a neural network for model adaptation. In \cite{weninger2019listen, meng2019speaker}, these adaptation methods were further investigated in end-to-end ASR \cite{chorowski2015attention, meng2019character}. However, all these methods focus on addressing the overfitting issue given very limited adaptation data in the target-domain.

Teacher-student (T/S) learning \cite{li2014learning, li2017large, meng2019conditional, meng2019domain} has shown to be effective for large-scale unsupervised domain adaptation by
minimizing the Kullback-Leibler
(KL) divergence between the output distributions of the teacher and student
models. 
The input to the teacher and
student models needs to be \emph{parallel} source- and target-domain
adaptation data, respectively, since the output vectors of a teacher network need to be frame-by-frame aligned with those of the student network to construct the KL divergence between two distributions.
Compared to one-hot labels, the
use of frame-level senone (tri-phone states) posteriors from the teacher as the soft targets to train the
student model well preserves the relationships among
different senones at the output of the teacher network. However, the parallel data
constraint of T/S learning restricts its application to
the scenario where the \emph{paired} target-domain data can be easily simulated
from the source domain data (e.g. from clean to noisy speech).
Actually, in many scenarios, the generation of parallel data in a new domain
is almost impossible, e.g., to simulate paired accented or
kids' speech from standard adults' speech.

Recently, adversarial learning \cite{gan, grl_ganin} was proposed for domain-invariant training \cite{grl_shinohara,  meng2018speaker, meng2018adversarial}, speaker adaptation \cite{meng2019asa}, speech enhancement \cite{pascual2017segan, meng2018cycle, meng2018afm} and speaker verification \cite{wang2018unsupervised, meng2019asv}. It was also shown to be effective for unsupervised domain
adaptation \emph{without} using parallel data \cite{grl_sun,dsn_meng}. In adversarial learning, an
auxiliary domain classifier is jointly optimized with the source model to
mini-maximize an adversarial loss. A deep representation is learned to be
invariant to domain shifts and discriminative to senone classification.
However, adversarial learning does not make use of the target-domain labels
which carry important class identity information and is only suitable for
the situation where neither parallel data nor target-domain labels are 
available. 

How to perform effective domain adaptation
using \emph{unpaired} source- and target-domain data with labels? We propose a neural label embedding (NLE) method:
instead of frame-by-frame knowledge transfer in T/S learning, we distill the knowledge of a source-domain model to a fixed set
of \emph{label embeddings}, or \emph{l-vectors}, one for each senone
class, and then transfer the knowledge to the target-domain model via these
senone-specific \emph{l}-vectors.
Each $l$-vector is a condensed representation of the DNN output
distributions given all the features aligned with the same senone at
the input. A simple DNN-based method is proposed to learn the $l$-vectors by minimizing the average $L_2$, Kullback-Leibler (KL) and
symmetric KL distance to the output vectors with the same senone label.
During adaptation, the \emph{l}-vectors are used in lieu of their corresponding one-hot labels to train the target-domain model with  cross-entropy loss.

NLE can be viewed as a knowledge quantization \cite{gray1984vector} in form of output-distribution vectors where each $l$-vector is a code-vector (centroid) corresponding to a senone codeword. With the NLE method, knowledge is transferred from the source-domain model to the
target-domain through a fixed codebook of \emph{senone-specific} \emph{l}-vectors instead of
variable-length \emph{frame-specific} output-distribution vectors in T/S learning.
These distilled \emph{l}-vectors decouple the target-domain model's output
distributions from those of the source-domain model and thus enable a more flexible
and efficient \emph{senone-level} knowledge transfer using \emph{unpaired} data.
When parallel data is available, compared to the T/S learning, NLE significantly reduces the computational cost
during adaptation by replacing the forward-propagation of each source-domain frame through the
source-domain model with a fast look-up in $l$-vector codebook. In the experiments, we adapt a
multi-conditional acoustic model trained with 6400 hours of US English to
each of the 9 different accented English (120 hours to 830 hours) and kids' speech (80
hours), the proposed NLE method achieves 5.4\% to 14.1\% and 6.0\% relative word
error rate (WER) reduction over one-hot label baseline on 9 accented English and kids' speech, respectively.

\section{Neural Label Embedding (NLE) for Domain Adaptation}


In this section, we present the NLE method for domain adaptation without using parallel data.
Initially, we have a well-trained source-domain network $M^S$ with parameters $\theta^S$ predicting
a set of senones $\mathbbm{C}$ and
source-domain speech frames $\mathbf{X}^S=\{\mathbf{x}^S_1, \ldots,
\mathbf{x}^S_{N_S}\}$ with senone labels $\mathbf{Y}^S=\{y^S_1, \ldots,
y^S_{N_S}\}$. We distill the knowledge of this
powerful source-domain model into a dictionary of \emph{l}-vectors, one for
each senone label (class) predicted at the output layer. Each \emph{l}-vector has the same
dimentionality as the number of senone classes. 
Before training the target-domain model $M^T$ with parameters $\theta^T$, we query the
dictionary with the ground-truth one-hot senone labels $\mathbf{Y}^T=\{y^T_1, \ldots, y^T_{N_T}\}$ of the target-domain speech frames $\mathbf{X}^T=\{\mathbf{x}^T_1, \ldots,
\mathbf{x}^T_{N_T}\}$ to get their corresponding \emph{l}-vectors. During
adaptation, in place of the one-hot labels, the
\emph{l}-vectors are used as the soft targets to train the target-domain 
model. For NLE domain adaptation, the source-domain data
$\mathbf{X}^S$ does not have to be parallel to the target-domain speech
frames $\mathbf{X}^T$, i.e., $\mathbf{X}^S$ and $\mathbf{X}^T$ do not have to be frame-by-frame synchronized and the number of frames $N_S$ does not have to be equal to $N_T$.

The key step of the NLE method is to learn \emph{l}-vectors from the
source-domain model and data. 
As the carrier of knowledge
transferred from the source-domain to the target-domain, the \emph{l}-vector $\mathbf{e}_c$
of a senone class $c$ should be a representation of the output distributions
(senone-posterior distributions) of the source-domain
DNN given features aligned with senone $c$ at the input, encoding the dependency between senone $c$ and all the other senones $\mathbbm{C}\setminus c$.  
A reasonable candidate is the \emph{centroid} vector that
minimizes the average distance to the output vectors generated from
all the frames aligned with senone $c$.
Therefore, we need to learn a dictionary of $|\mathbbm{C}|$ $l$-vectors corresponding to
$|\mathbbm{C}|$ senones in the complete set $\mathbbm{C}$, with each $l$-vector being $|\mathbbm{C}|$-dimensional.
To serve as the training target of the target-domain model, the $l$-vector $\mathbf{e}_{c}$ needs to be normalized such that its elements satisfy
\begin{align}
    e_{c,i} > 0, \quad \sum_{i=1}^{|\mathbbm{C}|} e_{c, i} = 1.
    \label{eqn:norm}
\end{align}


\subsection{NLE Based on $L_2$ Distance Minimization (NLE-L2)}
\label{sec:nle_l2}



To compute the senone-specific centroid, the most intuitive solution is to
minimize the average $L_2$ distance between the centroid and all the output
vectors with the same senone label, which is equivalent to calculating the arithmetic mean of the
output vectors aligned with the senone. Let $\mathbf{o}^S_n$ denote a $|\mathbbm{C}|$-dimensional output vector 
of $M^S$ given the input frame $\mathbf{x}^S_n$. $o^S_{n,i}$ equals to the
posterior probability of senone $i$ given $\mathbf{x}^S_n$, i.e., $o^S_{n,i} = P(i |
\mathbf{x}^S_n; \mathbf{\theta}^S), i \in \mathbbm{C}$.
For senone $c$, the \emph{l}-vector $\mathbf{e}_c$
based on $L_2$ distance minimization is computed as
\begin{align}
	\mathbf{\hat{e}}_{c} = \frac{1}{N_{S,c}} \sum_{n = 1}^{N_S} \mathbf{o}^S_{n}
	\mathbbm{1}[\mathbf{x}^S_n \in \text{senone} \; c], \;\; c \in
	\mathbbm{C},
	\label{eqn:nle_l2}
\end{align}
where $N_{S,c}$ is the number of source-domain frames aligned with senone $c$
and $N_S = \sum_{c\in \mathbbm{C}} N_{S,c}$.  The \emph{l}-vectors under
NLE-L2 are automatically normalized since each posterior vector $\mathbf{o}^S_{n}$ in the mean
computation satisfy Eq.~\eqref{eqn:norm}.  

\subsection{NLE Based on KL Distance Minimization (NLE-KL)}
\label{sec:nle_kl}
KL divergence is an effective metric to measure the distance between
two distributions. In NLE framework, the \emph{l}-vector $\mathbf{e}_c$ can
be learned as a centroid with a minimum average KL distance to the output
vectors of senone $c$. Many methods have been proposed to iteratively
compute the centroid of KL distance \cite{chaudhuri2008finding, veldhuis2002centroid, das2015kl}. 

\begin{figure}[htpb!]
	\centering
	\includegraphics[width=0.8\columnwidth]{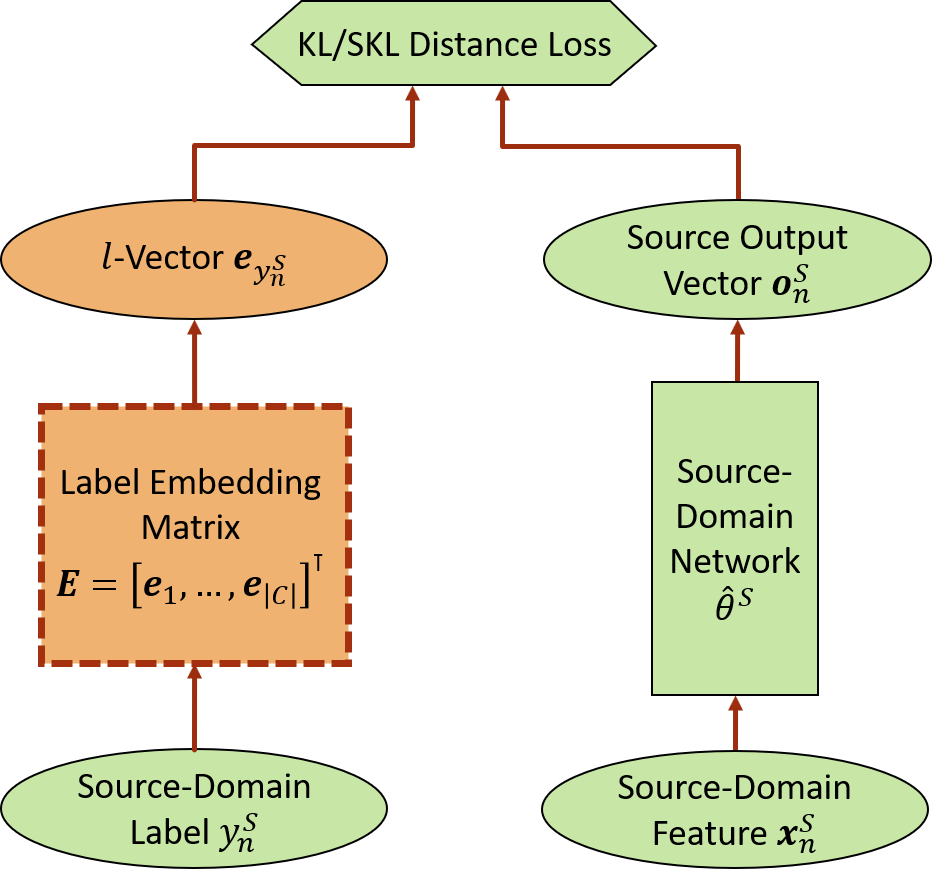}
    \vspace{-2pt}
    \caption{The diagram of learning neural label embeddings ($l$-vectors) through KL or SKL
    minimization. Only modules with red dotted lines are updated. }
	\label{fig:label_embed}
\end{figure}

In this paper, we propose a
simple DNN-based solution to compute this KL-based centroid. 
As shown in Fig. \ref{fig:label_embed}, we have an initial $|\mathbbm{C}|\times|\mathbbm{C}|$ embedding matrix $\mathbf{E}$ consisting
of all the $l$-vectors, i.e., $\mathbf{E}=[\mathbf{e}_{1}, \ldots,
\mathbf{e}_{|\mathbbm{C}|}]^{\top}$. For each source-domain sample, we look up the senone label
$y^S_n$ in $\mathbf{E}$ to get its $l$-vector $\mathbf{e}_{y^S_n}$
and forward-propagate $\mathbf{x}^S_n$ through $M^S$ to obtain the
output vector $\mathbf{o}^S_n$. The KL distance between
$\mathbf{o}^S_n$ and its corresponding centroid $l$-vector
$\mathbf{e}_{y^S_n}$ is 
\begin{align}
	\mathcal{KL}(\mathbf{e}_{y^S_n} || \mathbf{o}^S_{n}) = \sum_{i=1}^{|\mathbbm{C}|}
	e_{y^S_n, i} \log \frac{e_{y^S_n, i}}{o^S_{n, i}}.
\end{align}
We sum up all the KL distances and get the KL distance loss below
\begin{align}
	\mathcal{L}_{\text{NLE-KL}}(\mathbf{E}) = \frac{1}{N_S}\sum_{n=1}^{N_S} \mathcal{KL}(\mathbf{e}_{y^S_n} || \mathbf{o}^S_{n}). \label{eqn:loss_nle_kl}
\end{align}
To ensure each $l$-vector is normalized to satisfy Eq.~\eqref{eqn:norm}, we perform a softmax operation over a logit vector $\mathbf{z}_c \in \mathbbm{R}^{|\mathbbm{C}|}$ to obtain $\mathbf{e}_c$ below 
\begin{align}
    e_{c,i} = \frac{\exp(z_{c, i})}{\sum_{j=1}^{|\mathbbm{C}|} \exp{(z_{c, j})}}, \quad c \in \mathbbm{C}.
\end{align}
For fast convergence, $\mathbf{z}_c$ is initialized with the arithmetic mean of the pre-softmax logit vectors of the source-domain network aligned with senone $c$. The embedding matrix $\mathbf{E}$ is trained to minimize
$\mathcal{L}_{\text{NLE-KL}}(\mathbf{E})$ by updating 
$\mathbf{z}_1, \ldots, \mathbf{z}_{|\mathbbm{C}|}$
through standard back-propagation while the parameters $\mathbf{\theta}^S$ of $M^S$
are fixed. 

\subsection{NLE Based on Symmetric KL Distance Minimization (NLE-SKL)}
\label{sec:nle_skl}
One shortcoming of KL distance is that it is asymmetric: the minimization of
$\mathcal{KL}(\mathbf{e}_{y^S_n} || \mathbf{o}^S_{n})$ does not guarantee
$\mathcal{KL}(\mathbf{o}^S_{n} || \mathbf{e}_{y^S_n})$
is also minimized. SKL compensates for this by adding up the 
two KL terms together and is thus a more robust distance metric
for clustering. 
Therefore, for each senone, we learn a centroid $l$-vector with a
minimum average SKL distance to the output vectors of $M^S$ aligned with that senone 
by following the same DNN-based method in Section \ref{sec:nle_kl} except for replacing the KL distance loss with an SKL one. 

The SKL distance between an $l$-vector
$\mathbf{e}_{y^S_n}$ and an output vector $\mathbf{o}_n$ is defined as
\begin{align}
	 \mathcal{SKL}(\mathbf{e}_{y^S_n} || \mathbf{o}^S_{n}) = \sum_{i=1}^{|\mathbbm{C}|}
	\left( e_{y^S_n, i} - o^S_{n, i} \right) \log \frac{e_{y^S_n,
	i}}{o^S_{n, i}},
\end{align}
and the SKL distance loss is computed by summing up all pairs of SKL distances
between output vectors and their centroids as follows
\begin{align}
	\mathcal{L}_{\text{NLE-SKL}}(\mathbf{E}) = \frac{1}{N_S} \sum_{n=1}^{N_S} \mathcal{SKL}(\mathbf{e}_{y^S_n} || \mathbf{o}^S_{n}).
	\label{eqn:loss_nle_skl}
\end{align}

\subsection{Train Target-Domain Model with NLE}
\label{sec:train_kl}
As the condensed knowledge distilled from a large amount of source-domain
data, the $l$-vectors serve at the soft targets for training the target-domain model $M^T$.
\begin{figure}[htpb!]
	\centering
	\includegraphics[width=0.8\columnwidth]{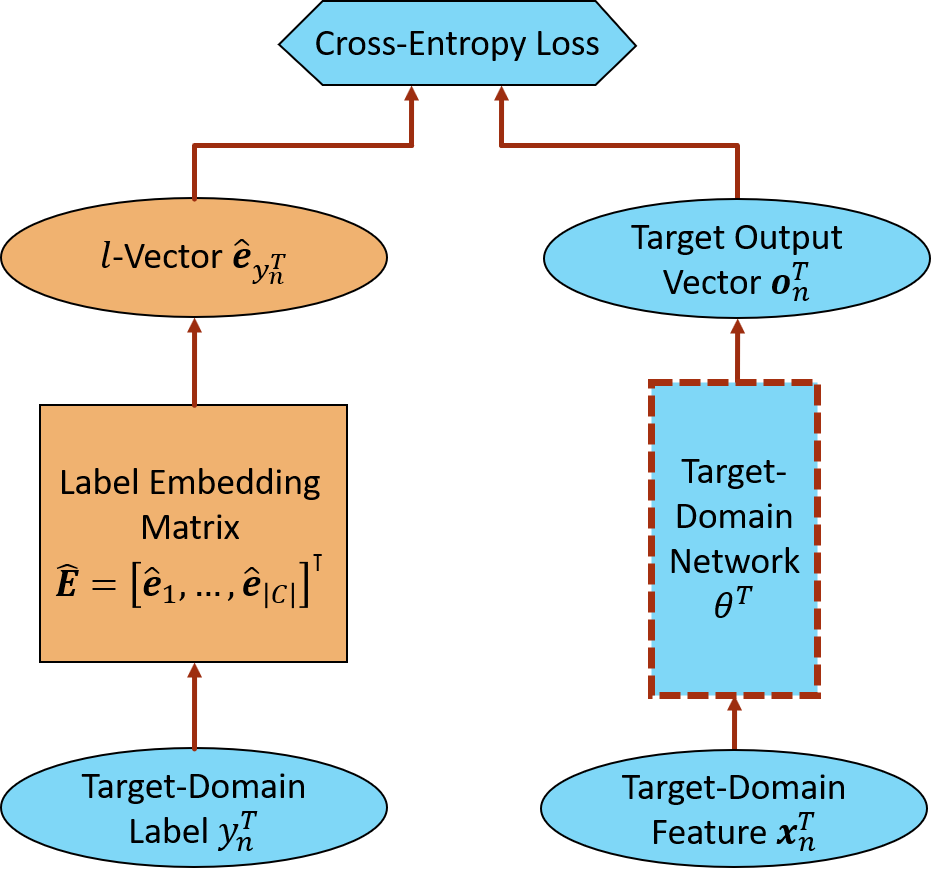}
    \vspace{-2pt}
    \caption{Train target-domain model using label embeddings
    ($l$-vectors). Only modules with red dotted lines are updated.}
	\label{fig:train_model}
\end{figure}
As shown in Fig. \ref{fig:train_model}, we look
up target-domain label $y^T_n$ in the optimized label embedding matrix $\mathbf{\hat{E}}$ for its $l$-vector $\mathbf{\hat{e}}_{y^T_n}$ 
 and forward-propagate $\mathbf{x}^T_n$ through $M^T$ to get
the output vector $\mathbf{o}^T_n$. 
We construct a cross-entropy loss using $l$-vectors $\mathbf{\hat{e}}^T_{y_n}$ as the soft targets below
\begin{align}
	\mathcal{L}_{\text{CE}}(\mathbf{\theta}^T) 
	& = \frac{1}{N_T} \sum_{n=1}^{N_T} \sum_{i=1}^{|\mathbbm{C}|}
	\hat{e}_{y^T_n, i} \log o^T_{n, i}, \label{eqn:loss_train_kl}
\end{align}
where $o^T_{n, i} = P(i| \mathbf{x}^T_n, \mathbf{\theta}^T), i \in \mathbbm{C}$ is the posterior of senone $i$ given $\mathbf{x}^T_n$.
We train $M^T$ to minimize
$\mathcal{L}_{\text{CE}}(\mathbf{\theta}^T)$ by updating only $\mathbf{\theta}^T$. 
The optimized $M^T$ with
$\hat{\theta}^T$ is used for decoding.

Compared with the traditional one-hot training targets that convey only
class identities, the soft $l$-vectors transfer additional quantized knowledge that
encodes the probabilistic relationships among different senone classes.
Benefiting from this, the NLE-adapted acoustic model is expected to achieve higher 
ASR performance than using one-hot labels on target-domain test data. The steps of NLE for domain adaptation are summarized in Algorithm
\ref{alg:nle}.

\begin{algorithm}
	\caption{Neural Label Embedding (NLE) for Domain Adaptation}
\label{alg:nle}
\begin{algorithmic}[1]
	\REQUIRE Source-domain model $M^S$, data $\mathbf{X}^S$, and 
	labels $\mathbf{Y}^S$. \\
	\quad\;  Target-domain data $\mathbf{X}^T$ and labels $\mathbf{Y}^T$. 
        \ENSURE Target-domain model $M^T$ with parameters $\hat{\theta}^T$ 
	
	\STATE \label{step:forward_src} Forward-propagate $\mathbf{X}^S$ through $M^S$ to generate
	output vectors $\mathbf{O}^S$. 
	\STATE Learn label embedding matrix 
	$\mathbf{E}_{L_2}$ by computing \emph{senone-specific} arithmetic means of $\mathbf{O}^S$ as in
	Eq.~\eqref{eqn:nle_l2}. 
	\REPEAT
	\STATE Forward-propagate $\mathbf{X}^S$ through $M^S$ to generate $\mathbf{O}^S$. 
	\STATE \label{step:nle_kl} Look up each $y^S_n$ in $\mathbf{E}_{\text{KL}}$  or $\mathbf{E}_{\text{SKL}}$ for
	$l$-vector $\mathbf{e}_{y^S_n}$. 
	\STATE Compute and
	back-propagate the error signal of loss $\mathcal{L}_{\text{NLE-KL}}$
	in Eq. \eqref{eqn:loss_nle_kl} or $\mathcal{L}_{\text{NLE-SKL}}$ in Eq. \eqref{eqn:loss_nle_skl} by updating $\mathbf{E}_{\text{KL}}$ or $\mathbf{E}_{\text{SKL}}$, respectively. 
	\UNTIL{convergence}
	
	\REPEAT \STATE \label{step:forward_tgt} Forward-propagate
	$\mathbf{X}^T$ through 
	$M^T$ to generate
	output vectors $\mathbf{O}^T$.
	\STATE \label{step:train_kl} Look up each $y^T_n$ in $\mathbf{\hat{E}}_{L_2}$, $\mathbf{\hat{E}}_{\text{KL}}$ or $\mathbf{\hat{E}}_{\text{KL}}$ for
	$l$-vector $\hat{\mathbf{e}}_{y^T_n}$. 
	\STATE Compute and
	back-propagate the error signal of loss
	$\mathcal{L}_{\text{CE}}$ in Eq.~\eqref{eqn:loss_train_kl} 
	by updating $\mathbf{\theta}^T$. 
	\UNTIL{convergence}
\end{algorithmic}
\end{algorithm}

\begin{table*}[!t]
\centering
\begin{tabular}[c]{C{1.0cm}|C{1.0cm}|C{1.0cm}|C{1.0cm}|C{1.0cm}|C{1.0cm}|C{1.0cm}|C{1.0cm}|C{1.0cm}|C{1.0cm}||C{1.0cm}}
	\hline
	\hline
	Task & A1 & A2 & A3 & A4 & A5 & A6 &
	A7 & A8 & A9 & Kids \\
	\hline
	Adapt & 160 & 140 & 190 & 120 & 150 & 830 & 250 & 330 & 150 & 80 \\
	\hline
	Test & 11 & 8 & 11 & 7 & 11 & 11 & 11 & 11 & 13 & 3 \\
	\hline
	\hline
\end{tabular}
\vspace{-5pt}
	\caption{Durations (hours) of adaptation and test data for each of
		the 9 accented English (A1-A9) and kids' speech.}
\label{table:data}
\end{table*}

\begin{table*}[!t]
\vspace{-3pt}
\centering
\begin{tabular}[c]{C{1.5cm}|C{1.0cm}
	|C{1.0cm}|C{1.0cm}|C{1.0cm}|C{1.0cm}|C{1.0cm}|C{1.0cm}|C{1.0cm}|C{1.0cm}||C{1.0cm}}
	\hline
	\hline
	Adapt. Method & A1 & A2 & A3 & A4 & A5 & A6 &
	A7 & A8 & A9 & Kids \\
	\hline
        Unadapted & 26.98 & 15.27 & 25.23 & 34.62 & 22.48 & 14.65 &	13.19  &
	14.91 & 9.80	&  27.83 \\
	\hline
        One-Hot & 20.37 & 14.46 & 20.14 & 19.99 & 15.06 &	12.50  & 11.73 	&
	13.90 & 9.71 & 26.99 \\ 
	\hline
        NLE-L2 & 18.39 & 12.91 & 18.54 & 18.39 & 14.14 &	12.04 &	10.54 &
	12.55 & 9.48 & 25.93 \\
	\hline
        NLE-KL & 18.30  & 12.86 & 18.74 & 18.42 & 14.25 &	11.90  & 10.39 &
	12.52 & 9.43 & 25.83 \\
	\hline
	NLE-SKL & \textbf{17.97} & \textbf{12.42} & \textbf{17.82} & \textbf{17.94}
	& \textbf{13.81} & \textbf{11.56} &	\textbf{10.15} &
	\textbf{12.21} & \textbf{9.19} & \textbf{25.36} \\
	\hline
	\hline
\end{tabular}
\vspace{-3pt}
\caption{ASR WERs (\%) of adapting a multi-conditional BLSTM acoustic model
	trained with 6400 hours US English to each
	of the 9 accented English and kids' speech with one-hot label and  the proposed NLE methods. }
\label{table:dialect_wer}
\vspace{-5pt}
\end{table*}

\section{Experiments}
We perform two domain adaptation tasks where parallel
source- and target-domain data is not accessible through data simulation: 
1) adapt a US English acoustic model
to accented English from 9 areas of the world; 2) adapt the same
acoustic model to kids' speech.
In both tasks, the source-domain training data is 6400 hours of multi-conditional
Microsoft US English production data, including Cortana, xBox and
Conversation data. The data is collected from mostly adults from all over the  
US. It is a mixture
of close-talk and far-field utterances from a variety of devices. 

For the first task, the adaptation data consists of 9 different types of accented English A1-A9 in which A1, A2, A3, A8 are from Europe, A4, A5, A6 are from Asia, A7 is from Oceania, A9 is from North America. A7-A9 are native accents because they are from countries where most people use English as their first language.  On the contrary, A1-A6 are non-native accents. 
Each English accent forms a specific target domain. For the
second task, the adaptation data is 80 hours of US English speech collected from
kids. The durations of different adaptation and test data are listed in Table \ref{table:data}. The training and adaptation data is transcribed. All data is anonymized with personally identifiable information removed. 

\subsection{Baseline System}
We train a source-domain bi-directional long short-term memory (BLSTM)-hidden Markov model
acoustic model \cite{sak2014long, erdogan2016multi,meng2017deep} with 6400
hours of training data.  This teacher model has 6 hidden layers with 600 units
in each layer.
80-dimensional log Mel filterbank features are
extracted from training, adaptation and test data. The output layer has
9404 units representing 9404 senone labels. The BLSTM is trained to minimize the frame-level
cross-entropy criterion. There is no frame stacking or skipping.
A 5-gram LM is used for decoding with around 148M n-grams. 
Table \ref{table:dialect_wer} (Row 1) shows the WERs of the
multi-conditional BLSTM
on different accents. This well-trained source-domain model 
is used as the initialization for all the subsequent re-training and adaptation
experiments.

For accent adaptation, we train an accent-dependent BLSTM for each accented English  
using one-hot label with cross-entropy loss. Each accent-dependent model is
trained with the speech of only one accent. As shown in Table
\ref{table:dialect_wer}, the one-hot re-training achieves 9.71\% to 20.37\%
WERs on different accents. For kids adaptation, we train a kids-dependent BLSTM using 
kids' speech with one-hot labels. In Table
\ref{table:dialect_wer}, we see that one-hot re-training achieves 26.99\% WER on
kids test data. We use these results as the baseline. 

Note that, in this work, we do not compare NLE with KLD adaptation \cite{kld_yu} 
since the effectiveness of KLD regularization reduces as the adaptation data increases and it is normally used when the adaptation data is very small (10 min or less).

\vspace{-2pt}
\subsection{NLE for Accent Adaptation}
\label{sec:dialect_adapt}
It is hard to simulate parallel accented speech from US English. We adapt the
6400 hours BLSTM acoustic model to 9 different English
accents using NLE. 
We learn 9404-dimensional $l$-vectors using NLE-L2, NLE-KL, and NLE-SKL 
as described in Sections \ref{sec:nle_kl} and \ref{sec:nle_skl} with the source-domain data and acoustic model. 
These $l$-vectors are used as the soft targets to
train the accent-dependent models with cross-entropy loss as in Section \ref{sec:train_kl}.

As shown in Table \ref{table:dialect_wer}, NLE-L2, NLE-KL, and NLE-SKL
achieve 9.48\% to 18.54\%, 9.43\% to 18.74\%, and 9.19\% to 17.97\% WERs,
respectively, on different accents.  NLE-SKL performs the best among the
three NLE adaptation methods, with 11.8\%, 14.1\%, 11.5\%, 10.3\%, 8.3\%,
7.5\%, 13.5\%, 12.2\%, and 5.4\% relative WER reductions over the one-hot
label baseline on A1 to A9, respectively. 
NLE-SKL
consistently outperforms NLE-L2 and NLE-KL on all the accents, with up to
4.0\% and 4.9\% relative WER reductions over NLE-L2 and NLE-KL, respectively. 
The relative reductions for native and non-native accents are similar 
except for A9. NLE-KL performs slightly better than
NLE-L2 on 6 out of 9 accents, but slightly worse than NLE-L2 on the other
3. All the three NLE methods achieve much smaller relative WER
reductions (about 5\%) on A9 than the other accents (about 10\%).
This is reasonable because North American English is much more similar to the
source-domain US English than the other accents. The source-domain model is
not adapted much to the accent of the target-domain speech.

\vspace{-2pt}
\subsection{NLE for Kid Adaptation}
Parallel kids' speech cannot be obtained through data simulation either. We adapt
the 6400 hours BLSTM acoustic model to the collected real kids'
speech using NLE. We use the same $l$-vectors learned in
Section \ref{sec:dialect_adapt} as the soft targets
to train the kid-dependent BLSTM acoustic model by minimizing the cross-entropy
loss. As shown in Table \ref{table:dialect_wer}, NLE-L2, NLE-KL, and NLE-SKL
achieve 25.93\%, 25.83\%, and 25.36\% WERs on kids' test set, respectively. NLE-SKL
outperforms the other two NLE methods with a 6.0\% relative WER
reduction over the one-hot baseline. We find that NLE 
is more effective for accent adaptation than kids adaptation. One
possible reason is that a portion of kids
are at the age of teenagers 
whose speech is very similar
to that of the adults' in the 6400 hours source-domain data. Note that all the
kids speech is collected in US and no accent adaptation is involved.


%
\section{Conclusion}
We propose a novel neural label embedding method for domain adaptation. 
Each senone label is represented by an $l$-vector that minimizes the average $L_2$, KL or SKL distances to all the source-domain output vectors aligned with the same senone. $l$-vectors are learned through a simple average or a proposed DNN-based method.
During adaptation, $l$-vectors serve as the soft targets to train the target-domain model. 
Without parallel data constraint as in T/S learning, NLE is specially suited for the situation where paired target-domain data samples cannot be simulated from the source-domain ones. Given parallel data, NLE has significantly lower computational cost than T/S learning during adaptation since it replaces the DNN forward-propagation with a fast dictionary lookup.

We adapt a multi-conditional BLSTM acoustic model trained with 6400 hours US English to 9 different accented English and kids' speech. NLE achieves 5.4\% to 14.1\% and 6.0\% relative WER reductions over one-hot label baseline. NLE-SKL consistently outperforms NLE-L2 and NLE-KL on all adaptation tasks by up to relatively 4.0\% and 4.9\%, respectively. As a simple arithmetic mean, NLE-L2 performs similar to NLE-KL with dramatically reduced computational cost for $l$-vector learning. 

\vfill\pagebreak
\bibliographystyle{IEEEbib}
\bibliography{refs}

\end{document}